\documentclass[12pt]{iopart}

\expandafter\let\csname equation*\endcsname\relax
\expandafter\let\csname endequation*\endcsname\relax
\usepackage{amsmath}
\usepackage{amssymb}
\usepackage{bm}
\usepackage{booktabs}
\usepackage{cite}
\usepackage{colortbl}
\usepackage{dsfont}
\usepackage{etoolbox}
\usepackage{fancyhdr}            
\usepackage{graphicx}
\usepackage{mathtools} 
\usepackage{xcolor}
\usepackage{xspace}
\usepackage{dcolumn}
\usepackage{slashed}
\usepackage{subcaption}
\usepackage{tikz-cd}
\usepackage[normalem]{ulem}

\definecolor{nicegreen}{rgb}{0.07, 0.564, 0.04}

\makeatletter
\newcommand{\mainmatter}{%
  \setcounter{footnote}{0}%
  \patchcmd{\@makefntext}{\fnsymbol}{\arabic}{}{}%
  \patchcmd{\@thefnmark}{\fnsymbol}{\arabic}{}{}%
  \def\@makefnmark{\textsuperscript{\arabic{footnote}}}%
}
\makeatother

\newcommand\ie{\mbox{\textit{i.\,e.}}\xspace}
\newcommand\cf{\mbox{c. f.}\xspace}
\newcommand\eg{\mbox{e.\,g.}\xspace}

\newcommand{\hx}{\hat{x}}
\newcommand{\hp}{\hat{p}}

\newcommand{\hk}{\hat{k}}
\newcommand{\hmx}{\hat{\mathcal{X}}}
\newcommand{\hmp}{\hat{\mathcal{P}}}

\newcommand{\sumi}{\sum_{I=1}^N}
\newcommand{\sumj}{\sum_{J=1}^N}

\DeclarePairedDelimiter\braket{\langle}{\rangle}
\DeclarePairedDelimiterX\Braket[2]{\langle}{\rangle}{#1 \delimsize\vert #2}
\DeclareUnicodeCharacter{2009}{\,} 
\DeclareUnicodeCharacter{2212}{-}
\definecolor{Gray}{gray}{0.7}

\begin{document}

\title{30 years in: \emph{Quo vadis} generalized uncertainty principle?} 

\author{Pasquale Bosso}
\address{Dipartimento di Ingegneria Industriale, Universit\`a degli Studi di Salerno, Via Giovanni Paolo II, 132 I-84084 Fisciano (SA), Italy}
\address{INFN, Sezione di Napoli, Gruppo collegato di Salerno, Via Giovanni Paolo II, 132 I-84084 Fisciano (SA), Italy}
\ead{pbosso@unisa.it}

\author{Giuseppe Gaetano Luciano}
\address{Applied Physics Section of Environmental Science Department, Escola Polit\`ecnica Superior,  Universitat de Lleida, Av. Jaume
II, 69, 25001 Lleida, Spain}
\ead{giuseppegaetano.luciano@udl.cat}

\author{Luciano Petruzziello}
\address{Dipartimento di Ingegneria Industriale, Universit\`a degli Studi di Salerno, Via Giovanni Paolo II, 132 I-84084 Fisciano (SA), Italy}
\address{INFN, Sezione di Napoli, Gruppo collegato di Salerno, Via Giovanni Paolo II, 132 I-84084 Fisciano (SA), Italy}
\address{Institut f\"ur Theoretische Physik, Albert-Einstein-Allee 11, Universit\"at Ulm, 89069 Ulm, Germany}
\ead{lupetruzziello@unisa.it}

\author{Fabian Wagner}
\address{Dipartimento di Ingegneria Industriale, Universit\`a degli Studi di Salerno, Via Giovanni Paolo II, 132 I-84084 Fisciano (SA), Italy}
\ead{fwagner@unisa.it}


\begin{abstract}
According to a number of arguments in quantum gravity, both model-dependent and model-independent, Heisenberg's uncertainty principle is modified when approaching the Planck scale. This deformation is attributed to the existence of a minimal length. The ensuing models have found entry into the literature under the term Generalized Uncertainty Principle (GUP). In this work, we discuss several conceptual shortcomings of the underlying framework and critically review recent developments in the field. In particular, we touch upon the issues of relativistic and field theoretical generalizations, the classical limit and the application to composite systems. Furthermore, we comment on subtleties involving the use of heuristic arguments instead of explicit calculations. Finally, we present an extensive list of constraints on the model parameter $\beta$, classifying them on the basis of the degree of rigour in their derivation and reconsidering the ones subject to problems associated with composites. 
\end{abstract}


\tableofcontents

\mainmatter

\section{Introduction}

Quantum theory and General Relativity are the two main pillars of our current understanding of physics. Yet, despite long research efforts and the surge of promising candidates, consensus on a predominant approach towards a unifying theory is still lacking. Besides conceptual issues, the development of theories of quantum gravity (QG) is severely complicated by the seeming experimental elusiveness of the Planck scale, which is where QG effects are expected to be strong. In the absence of direct empirical guidance, tentative progress can be made by concentrating on common predictions of the existing QG models. In this vein, arguments from different perspectives, such as string theory~\cite{Amati:1987wq,Amati:1988tn,Gross:1987ar,Gross:1987kza,Konishi:1989wk,Yoneya:1989ai}, asymptotically safe quantum gravity~\cite{Lauscher:2005qz,Ferrero:2022hor} and loop quantum gravity~\cite{Rovelli:1994ge,Modesto:2008jz,Girelli:2012ju}, as well as \emph{gedankenexperiments} heuristically combining quantum theory and general relativity \cite{Bronstein:1936lsk,Mead:1964zz,Mead:1966zz,Garay:1994en,Ng:1993jb,Amelino-Camelia:1994suz,Adler:1999bu,Bambi:2007ty,Padmanabhan:1987au,Maggiore:1993rv,Scardigli:1999jh,Calmet:2004mp,Susskind:2005js}, indicate that the classical-to-quantum transition of gravity could be marked by the emergence of a minimal length of order of the Planck scale ($l_{\text{P}}\sim 10^{-35}\,\mathrm{m}$).
It is worth mentioning that, in theories with additional dimensions, the value of the gravitational constant can increase, which implies a larger Planck length, \ie, a larger expected minimal-length scale \cite{Arkani-Hamed:1998jmv,Antoniadis:1998ig,Arkani-Hamed:1998sfv,Hossenfelder:2003jz}.

For, by now, 30 years, minimal-length effects have customarily been introduced into quantum mechanics by modifying the Heisenberg Uncertainty Principle (HUP).
As a result, the uncertainty of the position operator develops a global minimum when approaching the Planck scale.
The resulting uncertainty relation is commonly referred to as Generalized Uncertainty Principle (GUP)~\cite{Maggiore:1993kv,Kempf:1994su,Bang:2006va,Galan:2007ns,Pedram:2011aa,Pedram:2011gw,Tedesco:2011iv,Mignemi:2011wh,Nozari:2012gd,Das:2012rv,Tkachuk:2013lta,Pramanik:2013zy,Balasubramanian:2014pba,Dey:2018bmr,Bosso:2018uus,Chung:2018btu,Chung:2019raj,Bosso:2020aqm,Gine:2021nhu,Bosso:2022vlz,Fadel:2021hnx,Bosso:2017hoq,Wagner:2022bcy,HerkenhoffGomes:2022bnh}.
As time has gone by, this effective scheme has proven to be fertile ground for phenomenological explorations on the interplay between quantum and gravitational features.
In fact, a variety of  applications have hitherto been considered in single-particle quantum mechanics~\cite{Kempf:1996fz,
Brau:1999uv,
Chang:2001kn,
Akhoury:2003kc,
Benczik:2005bh,
Nozari:2005it,
Nozari:2005mr,
Nozari:2005np,
Fityo:2005xaa,
Brau:2006ca,
Quesne:2006fs,
Zhao:2007as,
Das:2008kaa,
Hossain:2008zz,
Bouaziz:2008wxq,
Ghosh:2009hp,
Das:2009hs,
Ali:2009zq,
Nozari:2010qy,
Das:2010zf,Bouaziz:2010he,Ali:2010yn,Pedram:2010hx,Chargui:2010zz,Pedram:2010zz,Pedram:2011xj,Hassanabadi:2012as,Pedram:2012ui,Vahedi:2012ym,Pedram:2012np,Benzair:2012yna,Pedram:2012my,Pedram:2012dm,Pedram:2012zz,Pedram:2012ub,Hassanabadi:2013jga,Hassanabadi:2013mq,Bouaziz:2013ora,Moayedi:2013nba,Blado:2013cqb,Asghari:2013sda,Ching:2013ala,AntonacciOakes:2013qvs,Faruque:2014epa,Das:2014bba,Faizal:2014mba,Hassanabadi:2014oja,Dey:2015gda,Bosso:2016frs,Wang:2016nkw,Shababi:2016ufq,Rossi:2016hrx,Deb:2016psq,Bouaziz:2017ijz,Bosso:2017ndq,Todorinov:2018arx,Park:2020zom,Twagirayezu:2020chx,Petruzziello:2020een,Girdhar:2020kfl,Aghababaei:2021mbi,Luciano:2021cna,Bosso:2021koi,Gubitosi:2021hoy,Bosso:2022ogb}, 
black hole physics~\cite{Adler:2001vs,Chen:2002tu,Cavaglia:2003qk,Medved:2004yu,Cavaglia:2004jw,Myung:2006qr,Nouicer:2007jg,Park:2007az,Xiang:2009yq,Myung:2009va,Jizba:2009qf,Scardigli:2014qka,Scardigli:2016pjs,Custodio:2003jp,Setare:2004sr,Setare:2005sj,Nozari:2005ah,Nozari:2005gn,Scardigli:2006eb,Zhao:2006ri,Ko:2006mr,Xiang:2006ei,Nozari:2006dq,Nozari:2006rt,Nozari:2006vn,Nozari:2006bi,Kim:2007hf,Nouicer:2007pu,Arraut:2008hc,Farmany:2009zz,Dehghani:2009zze,Banerjee:2010sd,Setare:2010ct,Said:2011dg,Majumder:2011xg,Sabri:2012fi,Tawfik:2013uza,Xiang:2013sza,Majumder:2013nia,Gangopadhyay:2013ofa,Chen:2014bva,Faizal:2014tea,Tawfik:2015kga,FaragAli:2015boi,NasserTawfik:2015gwy,Gangopadhyay:2015zma,Anacleto:2015rlz,Anacleto:2015awa,Anacleto:2015mma,Anacleto:2015kca,Hammad:2015dka,Dehghani:2015zca,Sakalli:2016mnk,Lambiase:2017adh,Ong:2018syk,Ong:2018xna,Maluf:2018lyu,Maluf:2018ksj,Alonso-Serrano:2018mfo,Alonso-Serrano:2018ycq,Contreras:2018gpl,Buoninfante:2019fwr,Kanazawa:2019llj,Hassanabadi:2019eol,Alonso-Serrano:2020hpb,Du:2021hxb,Bosso:2022xnm,Nozari:2008gp,Majumder:2012rtc,Nozari:2012nf,Chen:2013pra,Chen:2013tha,Chen:2013ssa,Feng:2015jlj,Ovgun:2015jna,Li:2016mwq,Ovgun:2017hje,Gecim:2017nbh,Gecim:2017zid,Kanzi:2019gtu,Buoninfante:2020cqz,Buoninfante:2020guu},
Cosmology~\cite{Chen:2003bu,Chen:2004ft,Battisti:2007jd,Battisti:2007wv,Battisti:2007zg,Bina:2007wj,Vakili:2007yz,Vakili:2008tt,Vakili:2008zg,Zhu:2008cg,Battisti:2008xy,Lidsey:2009xz,Basilakos:2010vs,Kim:2010wc,Hossain:2010wy,Chemissany:2011nq,Majumder:2011hy,Majumder:2011eg,Majumder:2012ph,Ali:2013qza,Ali:2013ii,Jalalzadeh:2013zwa,Ali:2014hma,Awad:2014bta,Awad:2014nma,Paliathanasis:2015cza,Garattini:2015aca,Moussa:2015bsa,Ali:2015ola,Atazadeh:2016yeh,Salah:2016kre,Khodadi:2018scn,Kouwn:2018rmp,Scardigli:2018jlm,Bosso:2019ljf,Blasone:2019wad,Gusson:2020pgh,Giardino:2020myz,Giacomini:2020zmv,Paliathanasis:2021egx,Luciano:2021vkl,Moussa:2021gxb,Nenmeli:2021orl},
Astrophysics~\cite{Nozari:2006rs,Casadio:2009jc,Tkachuk:2012gyq,Ghosh:2013qra,Ahmadi:2014cga,Feng:2016tyt,Vagenas:2017fwa,Ong:2018zqn,Ong:2018nzk,Bosso:2018ckz,Neves:2019lio,Moradpour:2019wpj,Viaggiu:2020iee,El-Nabulsi:2020hvt,Jusufi:2020wmp,Mathew:2020wnx,Abac:2020drc,Belfaqih:2021jvu,Anacleto:2021qoe,Abac:2021jpf,Blanchette:2021vid,Das:2021lrb,Tamburini:2021inp,Carvalho:2021ajy} 
and statistical mechanics~\cite{Nozari:2006gg,Tawfik:2012he,Li:2009bh,Vagenas:2018pez,Wang:2010ct,Wang:2011iv,Rashidi:2015rro,KalyanaRama:2001xd,Nozari:2006au,Ali:2011ap,Vakili:2012qt,Abbasiyan-Motlaq:2014kqa,Ali:2014dfa,Mathew:2017drw,Vagenas:2019wzd,Shababi:2020evc,Hamil:2020ldl,Luciano:2021ndh,Moradpour:2021ymp,Luciano:2022ely,Jizba:2022icu}, among others. On a different note, the idea of analogue systems has been gaining traction recently \cite{Conti:2014jsa,Braidotti:2016ido,Iorio:2017vtw,Iorio:2020olc,Iorio:2022ave,Iorio:2023uct}.

Three decades into minimal-length model building, it is about time that a report on the status of the field and its challenges be given. This is why, in this short review we highlight some of the difficulties encountered along the way. Note that we do not aim at thoroughly introducing the subject of GUPs; for a self-contained overview, the interested reader may consult the reviews \cite{Hossenfelder:2012jw,Tawfik:2014zca}, or for a more recent account Sec. 3 of \cite{Wagner:2022bcy}. We would like to emphasize that our choice of topics is unavoidably idiosyncratic, and do not claim comprehensiveness. The main intent behind the present work lies in turning the spotlight on some open problems of the field and providing guidelines on how to properly address them. A definitive solution inevitably requires more effort, and goes beyond the scope of the present review. 

For instance, it has been recently shown \cite{Amelino-Camelia:2022roq} that the uncertainty relation should receive relativistic corrections which are very much reminiscent of the GUP -- similarly to QG deformed wave functions, the single-particle sector of the Fock space in quantum field theory (QFT) inevitably acquires a spread in configuration space. However, as we show in Sec. \ref{RvG}, the two effects can be distinguished.

Another issue that has attracted attention in the last years is the classical limit. As it has been shown in \cite{Casadio:2020rsj}, the appearance of Planck's constant, if taken at face value, renders the ensuing classical theory trivial. In Sec. \ref{sec:conslim}, we show that the same reasoning applied to the speed of light removes minimal-length effects in the nonrelativistic regime, \ie for customary applications of the GUP. This indicates that it may be more instructive to understand the Planck mass $m_\text{P}$ as fundamental, \ie to keep it constant while letting $\hbar$ tend to 0, which is known as the relative-locality limit \cite{Amelino-Camelia:2011dwc}. As a result, the classical counterpart to GUP-deformed quantum theory is nontrivial.

On a different note, the introduction of an absolute minimal-length scale alone does not unambiguously fix the ensuing corrections to the dynamics embodied by the Hamiltonian \cite{Bosso:2023sxr}. Instead, as we demonstrate in Sec. \ref{sec:SymHam}, the choice of Hamiltonian is riddled with ambiguities. We also explain how symmetry arguments may help to find theoretically more appealing dynamics. For instance, in the relativistic context, it is well-known that the introduction of an absolute length scale requires a deformation of Lorentz symmetries to retain a relativity principle \cite{Amelino-Camelia:2000stu,Magueijo:2001cr}. We highlight that this subtlety may trickle down to the nonrelativistic regime, where Galilean invariance would have to be deformed \cite{Bosso:2022rue}. Otherwise, Galilean/Lorentz invariance is explicitly broken including the severe consequences these effects entail. 

The resolution of this issue inevitably entails a relativistic generalization of the GUP, on which as of yet no consensus exists. As the nonrelativistic limit can be obtained from relativistic dynamics but not viceversa, this is a highly underdetermined problem. As a result, a number of mutually inconsistent approaches have been put forward. Similarly, there are different proposals towards QFTs with deformations of GUP-type. These distinct approaches are collected in Sec. \ref{Rext}.

Tackling quantum fields, in turn, requires the introduction of multiparticle states. Section \ref{sec:FundConst} is devoted to clarifying some misconceptions surrounding meso- and macroscopic objects in the context of the GUP. In fact, considering center-of-mass motion, QG effects diminish with the number of elementary constituents contained in an object. If this was not the case, we would see minimal-length deformations on the level of, say, soccer balls \cite{Hossenfelder:2014ifa}.

In Sec. \ref{sec:heurexp} we compare heuristic and explicit approaches to different GUP deformed problems, and find a remarkable disparity between the results. These examples should act as a cautionary tale that heuristic reasoning should only be applied with due care and, if possible, be complemented with precise calculations.

Finally, we collect and classify, to our knowledge, all existing bounds on the dimensionless model parameter of the quadratic GUP (usually denoted $\beta$) in Sec. \ref{boun}. In doing so, we account for the misconceptions in the context of multiparticle states stated above, and list other arising problems. Taking those caveats into consideration, we find that the most stringent constraint 
\begin{equation}
    \beta<10^{30},
\end{equation}
obtained studying the hydrogen atom in the context of Lorentz-Invariance Violations (LIV) \cite{Gomes:2022hva}.

\section{Generalized uncertainty principles}
\label{GUPInt}

While there are alternative ways of approaching limits to localization \cite{Kothawala:2013maa,Kothawala:2014tya,Padmanabhan:2015vma,Lake:2018zeg,Lake:2019nmn,Lake:2019oaz,Lake:2020rwc,Dabrowski:2020ixn,Petruzziello:2020wkd,Wagner:2021thc}, it is customary to incorporate the minimal length into single-particle quantum mechanics as a deformation of the Heisenberg algebra. The modification of the uncertainty relation behind the term ``GUP'' is then immediately implied by the Robertson-Schr\"odinger relation \cite{Schroedinger:1930awq,Robertson:1929zz}. The present Section is intended to shortly introduce the reader into the subject and settle the notation.

In $d$ dimensions, we define a GUP as a nonrelativistic, quantum mechanical model featuring a deformed Heisenberg algebra\footnote{While there are also anisotropic models \cite{Gomes:2022hva,Wagner:2022rjg,HerkenhoffGomes:2023mqu}, these are relatively new and have not been dealt with thoroughly in the literature.} of the form \cite{Wagner:2021bqz,Bosso:2022ogb}
\begin{align}
    [\hx^a,\hp_b]=i\left[f(\hp^2)\delta^a_b+\bar f (\hp^2)\frac{\hp^a\hp_b}{\hp^2}\right],&&[\hx^a,\hx^b]=2i\theta(\hp^2)\hx^{[b}\hp^{a]},&&[\hp_a,\hp_b]=0,\label{DefHeisAlg}
\end{align}
where we introduced the analytic functions of the magnitude of the momentum $f,$ $\bar{f}$ and $\theta.$ These are related by the Jacobi identity $[[\hx^a,\hx^b],\hp_c]=2[[\hx^{[a},\hp_c],\hx^{b]}]$ such that
\begin{equation}
    \theta=2f'-\frac{\bar f}{\hp^2}\left[1+2(\log f)'\hp^2\right].
\end{equation}
Given the deformed Heisenberg algebra \eqref{DefHeisAlg}, the Robertson-Schr\"odinger relation \cite{Robertson:1929zz,Schroedinger:1930awq} (in its simpler but weaker Robertson form) becomes
\begin{equation}
    \Delta \hx^a\Delta \hp_b\geq \frac{1}{2}\left|\braket{f}\delta^a_b+\left\langle\bar f\,\frac{\hp^a\hp_b}{\hp^2}\right\rangle\right|.
\end{equation}
Considering directions $a=b,$ we may rewrite this relation as
\begin{equation}
    \Delta \hx^a\geq \frac{1}{2\Delta \hp_a}\left|\braket{f}+\left\langle\bar f\,\frac{\hp_{a}^2}{\hp^2}\right\rangle\right|.
\end{equation}
Here, the index $a$ is not being summed over. We obtain a minimum to localizability if the right-hand-side of this relation is bounded from below by a positive constant $\ell,$ the minimal-length scale. Take for example the often-applied model \cite{Brau:1999uv,Chang:2001bm,Brau:2006ca}
\begin{align}
    f=1+\beta l_{\text{P}}^2\hp^2,&&\bar f=\beta'l_{\text{P}}^2\hp^2,\label{eqn:ExMod}
\end{align}
with the the Planck length $l_{\text{P}}^2$, as well as the dimensionless model parameters $\beta$ and $\beta',$ supposed to be of order $1$. If these parameters are positive, we obtain
\begin{equation}
    \Delta \hx^a \geq \frac{1}{2\Delta \hp_a}\left[1+l_{\text{P}}^2\left(\beta\braket{\hp^2}+\beta' \braket{\hp_a^2}\right)\right]
    \geq \frac{1+(\beta+\beta')l_{\text{P}}^2\Delta p_a^2}{2\Delta p_a}
    \geq \sqrt{\beta+\beta'}l_{\text{P}}=\ell,\label{eqn:SecondOrderModel}
\end{equation}
where the last inequality involves minimization with respect to $\Delta p_a.$

While for the example given in Eq. \eqref{eqn:ExMod} the existence of a minimal length could be verified by means of simple algebraic manipulations, this procedure cannot be generalized to other models.
Recently, some of the authors of the present review have devised a different method to check generic models for the presence of a minimal uncertainty in the position \cite{Bosso:2023sxr} (related considerations can be found in \cite{Abdelkhalek:2016nyn,Segreto:2022clx}).
As it provides a new viewing angle on the GUP which will prove useful below, we present this approach here.

Given the operator $\hx^a,$ we can construct the conjugate operator $\hk_a$ such that
\begin{equation}
    [\hx^{a},\hk_{a}] = i,\label{eqn:PosWaveNumCom}
\end{equation}
where again the index $a$ is not being summed over.
Throughout this paper, we differentiate between the operator $\hk_a$, here termed wave number, and the momentum operator $\hp_a.$
While the former is \emph{defined} to be the conjugate operator to the position, the latter is a function of it.
It is worth emphasizing that this terminology makes no reference to the wave number as a property of a wave.
The exact form of the commutator $[\hx^a,\hk_b]$ for $a\neq b$ is irrelevant for the argument; for more details see \cite{Bosso:2023sxr}. The existence of a minimal length $\ell$ then poses a constraint on the spectrum of the wave-number operator. In particular, the space of wave numbers has to be bounded in such a way that
\begin{equation}
    \text{spec}(\hk_a)=\left\{k_a,\prod_{b\neq a}\lim_{k_b\to 0}k_a\in\left[-\frac{\pi}{2\ell},\frac{\pi}{2\ell}\right]\right\}.\label{eqn:WaveNumSpec}
\end{equation}
For example, in one dimension this condition reduces to $\text{spec}(\hk)=\{k,k\in[-\pi/2\ell,\pi/2\ell]\}.$ Given a model as in Eq. \eqref{DefHeisAlg}, the operator $\hk_a$ can be obtained as a function of $\hp_a$ such that this constraint (and with it the presence of a minimal length) can be checked by analyzing the domain of $\hk_a(\hp_b).$
In turn, we can always define a wave number $\hat{k}_a$ satisfying Eq. \eqref{eqn:PosWaveNumCom}.
From this, we can define a momentum operator $\hat{p}_a$ satisfying the algebra in Eq. \eqref{DefHeisAlg}.
Thus, the momentum $\hat{p}_a$ is unnecessary in defining a minimal length.
In fact, the very definition of the momentum operator is ambiguous:
given the wave number $\hat{k}_a$ and the corresponding bound related to the minimal length, for any choice of the modifying functions in Eq. \eqref{DefHeisAlg} we obtain a specific momentum $\hat{p}_a$.
However, no specific choice of the modifying functions (and thus of the momentum) can be deemed as the correct one.
Therefore, the momentum operator $\hat{p}_a$ cannot \emph{de facto} be related to the underlying minimal length.

The presence of a minimal length, in turn, has operational implications for the techniques used in solving problems.
In particular, without the possibility of infinite localization, the position operator ceases to be self-adjoint (while staying symmetric) \cite{Kempf:1994su}, thereby making it impossible to define physical position eigenstates.
Instead, we can define a quasi-position representation \cite{Kempf:1994su,Detournay:2002fq,Bernardo:2018xmu,Gomes:2023yhj}.
Such a representation is constructed projecting states on the overcomplete set of states of minimal position uncertainty \cite{Bosso:2020aqm}.
To provide a physical model, the kinematics of the minimal-length deformation (represented by Eq. \eqref{DefHeisAlg}) has to be complemented by dynamics. It is customary to define the underlying single-particle Hamiltonian as
\begin{equation}
    \hat{H} = \frac{\delta^{ab}\hp_a\hp_b}{2m}+V(\hx^a),\label{eqn:GUPHam}
\end{equation}
where $m$ stands for the mass of the particle at hand. This Hamiltonian automatically implies universal corrections \cite{Das:2008kaa} to eigenvalues of observables as well as Heisenberg/Schrödinger dynamics (or Hamiltonian dynamics as classical counterpart), which allows for a rich phenomenology as indicated in the introduction. 

To put it in a nutshell, GUPs are a precisely defined class of models. Being precisely defined, however, does not preclude the existence of subtleties. This is the matter of the subsequent Section.

\section{Open problems}
\label{Op}

The present Section is dedicated to presenting some of the open problems and issues plaguing the field of GUPs. In particular, we compare relativistic and minimal-length-induced corrections to uncertainty relations. We further discuss ambiguities of the classical limit.
Thereupon, the particular choice of Hamiltonian, Eq. \eqref{eqn:GUPHam}, is scrutinized. This Hamiltonian is to emerge as the nonrelativistic limit of an underlying relativistic model. We review approaches towards this relativistic extension, pointing out the lack of consensus on the matter and touching upon QFT. In this context, the issue of multiparticle states, which has been the subject of misconceptions we highlight, has to be developed. Furthermore, we note that there are disparities in results obtained by heuristic and explicit approaches to minimal-length models, warning against the light use of the former. Finally, we present the aforementioned collection of bounds, thereby correcting some misunderstandings in the literature.

\subsection{Relativistic vs. GUP corrections}
\label{RvG}
It has recently been pointed out that relativistic corrections to quantum mechanics derived from the single-particle sector of scalar quantum field theory may lead to uncertainty relations of the form \cite{Amelino-Camelia:2022roq}
\begin{equation}
   \Delta x \Delta p \gtrsim \frac{\hbar}{2}\left(1+\frac{3\Delta p^2}{4m^2c^2}\right) =\frac{\hbar}{2}\left(1+\frac{ 3\lambda_C^2\Delta p^2}{4\hbar^2}\right),
\end{equation}
with the mass of the particle in question $m$ and its Compton wavelength $\lambda_C.$ The similarity to Eq. \eqref{eqn:SecondOrderModel} is obvious. Indeed, the effect even allows for an analogous interpretation to the GUP. In contrast to nonrelativistic positions, the eigenstates of the Newton-Wigner position operator \cite{Newton:1949cq} are not given by Dirac-delta functions, but have finite width. This is analogous to the quasi-position representation of the GUP \cite{Bosso:2020aqm}.

This observation begs the questions: in which way do GUP-corrections differ from relativistic ones?
and can GUP-effects sincerely be deemed nonrelativistic?
To approach these questions, let us expand the ordinary, special-relativistic dispersion relation in power series of $k/mc$
\begin{equation}
    E = \sqrt{k^2c^2+m^2c^4}\simeq mc^2+\frac{k^2}{2m}+\frac{k^4}{8m^3c^2}.
\end{equation} 
We thus obtain the relativistic correction to the Hamiltonian $\delta \hat{H}_{\text{Rel}} = \lambda_C^2\hat{k}^4/8\hbar^2m.$
The corresponding typical GUP-induced correction to the Hamiltonian reads $\delta \hat{H}_{\text{GUP}} = \beta\ell^2\hat{k}^4/\hbar^2m$.
Although mass-dependent models have also been proposed \cite{Maggiore:1993rv,Maggiore:1993kv,Fadel:2021hnx,Wagner:2023fmb}, these do not affect the current argument.
Clearly, the relativistic and GUP-corrections are very similar. 
However, while the Compton wavelength appearing in the relativistic model is particle-specific, the minimal length $\ell$ ought to be universal when applied to elementary particles (see Sec. \ref{sec:FundConst} and \cite{Quesne:2009vc,Tkachuk:2012gyq,Amelino-Camelia:2013fxa}).
Thus, both can be told apart once different particle species are considered.

Yet, this does not imply that the GUP can be understood to be of zeroth order in a $1/c$-expansion, \ie that it is nonrelativistic. This is the subject of the subsequent subsection.

\subsection{Consistent limits}
\label{sec:conslim}

In the GUP literature, it has been common practise for a long time (see for instance \cite{KalyanaRama:2001xd,Chang:2001bm,Benczik:2002tt}) to investigate aspects of classical dynamics with deformations of GUP-type. However, it has been shown in a recent study \cite{Casadio:2020rsj} that the classical limit of the GUP is more involved than just naïvely applying the transformation
\begin{equation}
    \frac{[\hx,\hp]}{i\hbar}\rightarrow \{x,p\}.
    \label{eqn:QMtoC}
\end{equation}
The issue lies in the fact that the length scale $\ell$ is understood to be proportional to the Planck length $l_{\text{P}}=\sqrt{\hbar G/c^3}.$ For reasons of clarity, here we resort to quadratic corrections to the Heisenberg algebra and work in one spatial dimension. As a result, the modified commutator in one dimension may be written as
\begin{equation}
    [\hx,\hp]=i\hbar\left(1+\beta \frac{G}{\hbar c^3}\hp^2\right),
\end{equation}
where $\beta$ again amounts to a dimensionless model parameter. Classical physics should be recovered in the limit of vanishing $\hbar$ on the level of expectation values
\begin{equation}
    \{x,p\}=\lim_{\hbar\rightarrow 0}\frac{\braket{[\hx,\hp]}}{i\hbar}=1+\frac{\beta G}{c^3}\lim_{\hbar\rightarrow 0}\frac{\braket{p^2}}{\hbar}.
\end{equation}
While the latter term depends on the quantum state the system is in, it has been convincingly shown that, due to the additional factor of $\hbar^{-1},$ the resulting Poisson brackets either diverge (a meaningless result) or stay unmodified \cite{Casadio:2020rsj}. Thus, it appears that GUP effects necessarily dissolve in the classical limit.\footnote{The classical and quantum dynamics of the GUP have also been contrasted in the context of Koopmann-von Neumann mechanics \cite{Chashchina:2019uyt}. As a result, the classical version of the theory necessarily acquires a quantum nature, a contradiction that appears to corroborate the reasoning in \cite{Casadio:2020rsj}. However, this finding rests crucially on a somewhat idiosyncratic choice of modification of the underlying algebra whose quantum mechanical counterpart does not reproduce the well-known GUP-deformed dynamics. After accounting for this peculiarity, the effect disappears.}

There are three important lessons to be drawn on the basis of this calculation:
\begin{itemize}
    \item First, the classical limit is not the only regime of interest in quantum gravity phenomenology. If we, instead, take the relative-locality limit $G\rightarrow 0,$ $\hbar\rightarrow 0$ with $G/\hbar=\text{const},$\footnote{In other words, while the Planck mass $m_\text{P}$ continues to be a relevant constant, the Planck length $l_{\text{P}}$ goes to zero in the limit.} the corrections survive, eventually leading to modified Poisson brackets. As the name suggests, this viewpoint has been the subject of a number of studies in the context of doubly/deformed special relativity and relative locality \cite{Amelino-Camelia:2011lvm}. Importantly, this makes it impossible to apply the resulting modifications to gravitational physics or dynamics on curved backgrounds of the kind studied, for instance, in
    \cite{Wang:2010ct,Tkachuk:2012gyq,Ghosh:2013qra,Vagenas:2017fwa}.
    \item Second, the corrections to the algebra are of the order $c^{-3},$ \ie relativistic.
    Therefore, they \emph{a priori} should vanish when considering nonrelativistic quantum mechanics.
    More precisely, in the $1/c$-expansion of the Hamiltonian, the modifying term should appear after the first relativistic corrections. 
    Thus, the same reasoning that trivializes GUP corrections to classical problems would also render them meaningless in the nonrelativistic regime.
    
    \item Third, neither is the speed of light generally infinite, nor do Planck's and Newton's constants vanish in reality. Taking limits in one of the three dimensionful constants without considering the effect of the other two can lead to faulty conclusions. For phenomenological purposes, it can be sensible to consider QG corrections to classical or nonrelativistic dynamics even though they may only appear at higher orders in perturbative expansions of dimensionful fundamental constants as long as the effect can be distinguished from other, possibly much larger contributions.
\end{itemize}
We thus observe the complication of considering limits in different quantities at the same time. It can be well-motivated to study corrections to classical dynamics from Planck-scale suppressed effects, even though the interpretation is subtler then at the quantum level. However, classical objects are usually macroscopic, thus complicating the application of GUPs (see Sec. \ref{sec:FundConst}).

Furthermore, due to them being intrinsically relativistic, it is important to try to understand the relativistic completion as well as the symmetries underlying GUP-deformed theories, an area of research which, as of yet, has not been adequately addressed.

\subsection{Symmetries and the choice of Hamiltonian}
\label{sec:SymHam}

In Sec. \ref{GUPInt}, we provided a short introduction into the field of GUPs by first defining a deformed commutator between $\hx^a$ and $\hp_b$ (\cf Eq. \eqref{DefHeisAlg}), and then providing a model Hamiltonian in Eq. \eqref{eqn:GUPHam}, which we repeat here for convenience 
\begin{equation}
    \hat{H}=\frac{\delta^{ab}\hp_a\hp_b}{2m}+V(\hx^a).
    \tag{\ref{eqn:GUPHam}}
\end{equation}
However, we also showed (\cf see Eq. \eqref{eqn:WaveNumSpec} and the ensuing comments below) that the mere existence of a minimal length does not fix the definition of the momentum $\hp_a.$
Given a model as in Eq. \eqref{DefHeisAlg}, we can check for a minimal length, but that minimal length is just a byproduct of the deformation. In other words, simply proposing the Hamiltonian in Eq. \eqref{eqn:GUPHam} corresponds to a \emph{top-down} approach. Yet, an exact intuition on a specific form of Eq. \eqref{DefHeisAlg} and posterior definition of Hamiltonian, as is usually required from top-down reasoning, is lacking.

Given the wave-number conjugate to the position $\hk_a$ introduced in Eq. \eqref{eqn:PosWaveNumCom}, there is a constructive statement that can be made from the sole presence of the minimal-length scale itself. The boundedness of the ensuing wave-number space is not only a tool to check for the existence of a minimal length, but it can also be used as basis for a \emph{bottom-up} treatment of the ensuing models. The existence of an operator $\hp_a$ satisfying Eq. \eqref{DefHeisAlg} is not required on the level of kinematics; it amounts to an additional structure whose definition is somewhat arbitrary. In other words, kinematically, there is only one model of minimal-length quantum mechanics.

The definition of a momentum $\hp_a,$ in turn, only has physical consequences once it is incorporated into the dynamics through the Hamiltonian given in Eq. \eqref{eqn:GUPHam}. However, its definition was arbitrary in the first place, and the Hamiltonian in Eq. \eqref{eqn:GUPHam} inherits this degree of arbitrariness. Why, for example should we not have chosen
\begin{equation}
    \hat{H}=\frac{\delta^{ab}\hk_a\hk_b}{2m}+V(\hx^a),    
\end{equation}
instead of Eq. \eqref{eqn:GUPHam}? why not any other function of $\hk_a$ and $\hx^b$ which reduces to the ordinary quantum mechanical Hamiltonian in the limit of vanishing minimal length? At first sight, there is no physical reason to choose one over the other.

What could thus be an informed guess on the form of the Hamiltonian? When constructing dynamics, it is usually helpful to revert to spacetime symmetries. It has been shown that the Hamiltonian given in Eq. \eqref{eqn:GUPHam} violates Galilean invariance \cite{Bosso:2022rue}. 
This can be proved by studying the algebra of operators under the effects of space and time transformations, without no reference to the classical counterpart.
In particular, it is not subject to the subtleties of the classical limit (\cf Sec. \ref{sec:conslim}).
However, the consequences of this violation of Galilean invariance have not been worked out yet.

Be that as it may, a comprehensive analysis of the symmetries of Eq. \eqref{eqn:GUPHam} is still lacking. Taking inspiration from doubly special relativity, Galilean invariance may be deformed instead of being strictly broken. This property, in turn could provide constraints on the allowed forms of Hamiltonian. 

Relativistic models of deformed special relativity could then be understood as algebra deformations of GUP models introducing the speed of light as deformation parameter.
\footnote{Deformed special relativity would then be a twofold deformation of Galilean relativity with the deformation parameters $\ell$ and $c.$}
GUP models, in turn, could be derived from DSR by a nonrelativistic limit.\footnote{In the context of modified dispersion relations, this link has been found in \cite{Wagner:2023prep}.} This is one of the possible relativistic generalizations of the GUP which are the subject of the subsequent subsection. 

\subsection{Relativistic extension}
\label{Rext}
The Heisenberg algebra relates positions and momenta. While simple analogy with the nonrelativistic case appears to indicate a straightforward extension to Lorentzian spacetime of the form
\begin{equation}
[\hx^\mu,\hp_\nu]=i\hbar\delta^\mu_\nu,\label{RelHeisAlg}
\end{equation}
there are a number of shortcomings of this approach:
\begin{itemize}
    \item First, in both relativistic as well as nonrelativistic quantum mechanics time is a parameter, not an operator (for some recent literature on this subject consult \cite{Muga:2002kad} and references therein). In particular, as had been first pointed out by Pauli \cite{Pauli:1933akf} and rigorously shown in \cite{Srinivas:1981abc}, it is impossible to define a self-adjoint time operator for physical systems described by a Hamiltonian with spectrum bounded from below.
    \item Second, the Heisenberg algebra is a hallmark of single-particle quantum theory. After all, the position $x^\mu$ represents a single worldline. However, a consistent extension of a quantum theory to the relativistic realm clearly necessitates the possibility of particle creation. It intrinsically has to be a theory of many particles, \ie a QFT (for an instructive explanation of this fact see chapter one in \cite{Padmanabhan:2016xjk}). In those, positions, not being Poincaré-charges, are not described by operators (leaving aside the Newton-Wigner operator representing spatial positions in the single-particle sector only \cite{Newton:1949cq}). As a result, it is unclear how to interpret the deformation of an uncertainty relation of the type \eqref{RelHeisAlg}.
\end{itemize}

However, instead of concentrating on the commutation relations of physical operators, it may be sufficient to consider the non-canonical transformation towards canonical variables. In particular, while the interpretation of non-coincident physical and canonical positions is unclear in QFT, the particular choice of unequal canonical and physical momenta can be made sense of. In other words, as above we may define the canonical conjugates to the positions, the wave numbers
\begin{equation}
   \hk_\mu=\hk_\mu (\hp).
\end{equation}
Clearly, this procedure is only possible for GUPs which can be represented in this way, \ie the ones set on a commutative geometry (for more details see \cite{Bosso:2022ogb}).\footnote{Noncommutative geometries allow for a definition of wave numbers \cite{Bosso:2023sxr}. Those, however, cannot be represented by gradients because the noncommutativity of the coordinates precludes the existence of a position representation. In this case, other methods such as star-product realizations \cite{MELJANAC:2009doq,Battisti:2010sr,Meljanac:2011mt,Meljanac:2017qck} are required.}

Following the GUP philosophy, the relativistic dispersion relation may be understood in terms of the momentum $p_\mu(k),$ thus reading
\begin{equation}
    -p^2(k)=m^2,
\end{equation}
with $m$ the mass of the particle under investigation. This modification may be applied in a covariant way,\footnote{While this may appear trivial at first sight, at the level of many particles it motivates higher-derivative, possibly nonlocal quantum field theories \cite{Todorinov:2018arx}.} or break/deform Lorentz invariance \cite{Bosso:2022ogb}.

This disparity brings to the surface the fundamental ambiguity from which relativistic extensions of the GUP innately suffer. How should the time component of the physical momentum be expressed in terms of its canonical counterpart?
Say, for example, in the nonrelativistic regime momentum and wave number are related as $\hp_a=f(\hk^2)\hk_a.$ Two different relativistic generalisations immediately come to mind. On the one hand, if understood covariantly, a correction of the form $\hp_\mu=f(\hk_\mu\hk_\nu\eta^{\mu\nu})\hk_\mu$ is to be expected.\footnote{Note, however, that this modification does not recover the GUP-deformed Schr\"odinger equation in the nonrelativistic limit.} On the other hand, the energy may as well be chosen to stay unmodified such that $\hp_\mu=(\hk_0,f(\hk^2)\hk_a),$ thus breaking (or deforming) Lorentz symmetry. At best, we can thus say that a relativistic extension is inspired by an underlying GUP, but it cannot be derived unambiguously. 

On the conceptual level, it is important to bear in mind that these choices may have deeply differing implications. Do we, for example, want to consider a minimal length only or a minimal time as well? If these statements are to be covariant, could it be more helpful to consider a minimal spacetime volume?
\footnote{This covariant form of discreteness is the basis for the Causal-Set approach to QG \cite{Bombelli:1987aa,Rideout:1999ub}(see \cite{Sorkin:2003bx,Surya:2019ndm} for reviews).}

These questions are all the more important because at high energies a relativistic description is without alternative. The scale of the considered interaction (say, the center-of-mass energy in scattering processes), in turn, is the most obvious amplifier of Planck-scale effects. But a relativistic description itself will not do. To make consistent predictions, it is required to consider deformations of the QFTs comprising the standard model. 

Regarding QFTs, up until now, no consensus has been reached on the level of methods or techniques. The following approaches have been worked out in the past.
\begin{itemize}
    \item A first ansatz made use of the Bergmann-Fock formalism \cite{Kempf:1994fv,Kempf:1996ss}, while introducing a minimal length as well as a minimal momentum. This idea has the added advantage that it circumvents the issue that there are no position and momentum eigenstates in these backgrounds. Instead, the theory is defined in a deformed but regular Bargmann-Fock basis.
    
    \item The same problem has been tackled in the context of canonical quantization. Specifically, using standard techniques related to models with a minimal length, fields are first described in terms of the corresponding modes, thus represented in momentum space \cite{Nouicer:2005dp,Frassino:2011aa,Bosso:2021koi}.
    Hence, this approach reverts to modelling fields as a collection of an infinite number of harmonic oscillators, thereby closely resembling the classic route to fields.
     As such, it avoids the cumbersome use of the quasi-position representation.
    
    \item A possible strategy consists in retaining the ordinary form of the position-space field equations while replacing the properly modified operators \cite{Adler:1999js,Nozari:2005ix,Quesne:2006is,Kober:2010sj,Chargui:2010zz,Kober:2011dn,Moayedi:2011ur,Pedram:2011zz,Hassanabadi:2013jga,Moayedi:2013nba,Moayedi:2013nxa,Faizal:2014rwa,Bosso:2020jay}.
    However, this approach can at best be perturbatively connected to the GUP. Recall that there is no position representation in this context. As a result, this produces a higher-derivative theory, often raising the issues that such theories ordinarily present, such as Ostrogradsky instabilities and the presence of ghost fields.
    \item Alternatively, it is possible to apply an adequate generalization of the Fourier transform \cite{Bosso:2020aqm} from momentum space to the maximal-localization states. This ansatz grants a consistent quasi-position representation of the fields and the corresponding equations of motion \cite{Matsuo:2005fb}. 
    \item A further alternative is constituted by a path-integral approach, in which a momentum-space description of the Lagrangian density represents the obvious starting ground \cite{Kempf:1995zn,Matsuo:2005fb,Benzair:2012yna}.
    Once again, since the density is then integrated over all possible values of the momentum, special care must be dedicated to the ultraviolet sector of the model.
    \item Several authors have instead preferred a completely different approach: that of imposing specific commutation relations between the field operators and their conjugates or between the field operators and their canonical momenta \cite{Kober:2011uj,Husain:2012im}.
    Typically, such modified commutators are inspired by the position-momentum commutators of nonrelativistic quantum mechanics \eqref{DefHeisAlg}, \emph{i.e.}, considering modifications depending on the canonical field momentum.
    However, in this approach, it is not always clear whether a minimal length is present.
\end{itemize}
Summarizing, there are a number of partially mutually inconsistent approaches towards QFT with a minimal length. A comprehensive comparative study of these different ideas, possibly unearthing connections between them, is lacking. Therefore, it is difficult to assess the state of the field regarding relativistic extensions.

In order to be able to study QFT with a certain degree of rigour, it is necessary to understand multiparticle states, which are subject of the subsequent subsection.

\subsection{Fundamental constituents}
\label{sec:FundConst}

A consistent theory of mechanics, which the GUP ought to be, must provide a description of composite particles.
In typical analyses involving nonrelativistic quantum mechanics in the mesoscopic regime, the GUP is applied to both elementary particles and macroscopic aggregates (such as large molecular compounds) in exactly the same way. However, there are serious arguments against such a na\"ive application to macroscopic objects \cite{Amelino-Camelia:2013fxa}.

In the framework of doubly special relativity \cite{Amelino-Camelia:2000stu,Magueijo:2001cr}, loop quantum gravity \cite{Bojowald:2012ux,Assanioussi:2014xmz,Brahma:2018rrg} as well as three-dimensional gravity coupled to matter \cite{Freidel:2005me,Freidel:2005bb}, it has been argued that the composition law for momenta needs to be deformed such that it becomes nonlinear. Such a deformation can always be treated as a small perturbation as long as the characteristic physical quantities of the considered system are not comparable with the Planck scale. Hence, this implies the absence of inconsistencies in the study of elementary particles. However, the above requirement breaks down when considering composite objects made up of a large number of small constituents, for which, e.g., the Planck mass ($\simeq 10^{-5}$ g) does not represent an out-of-reach threshold. Therefore, it appears that, for such macroscopic systems, the nonlinear terms of the composition law of momenta should play a relevant role, but experimentally this is not the case. Since in our macroscopic world there is no imprint of QG, the said nonlinearity turns out to be an undesirable feature of the model. This issue is typically referred to as the ``soccer-ball problem'' \cite{Amelino-Camelia:2014gga,Amelino-Camelia:2003ezw,Hossenfelder:2014ifa}, and it has been addressed with several proposals \cite{Hossenfelder:2007fy,Amelino-Camelia:2011dwc,Amelino-Camelia:2014gga,Amelino-Camelia:2020vvl}.

It may be tempting to transfer the reasoning developed so far in the context of the deformed composition of momenta and argue that the GUP is affected by the same ill-defined macroscopic limit. However, a simple translation of this conclusion, while highlighting a related issue, is not possible. On the contrary, QG effects deteriorate with increasing number of elementary constituents of a composite system rather than being enhanced.
Although this conclusion constitutes a serious obstacle to the experimental accessibility of QG effects, a similar scaling may also be advocated to compare bounds from with available data, especially in the astrophysical/cosmological scenario (see in particular the comment in Ref. \cite{Quesne:2009vc} on Ref. \cite{Benczik:2002tt}.)

To show this argument in a simple scenario, we extend the reasoning already proposed in \cite{Quesne:2009vc,Tkachuk:2012gyq,Amelino-Camelia:2013fxa} to a general class of deformed uncertainty relations. The starting point consists in introducing the modified commutation relations for the position and momentum operators belonging to the elements of an $N$-particle composite system, that is 
\begin{align}
    \left[\hx^i_I,\hp_{J,j}\right] =& i\delta_{IJ}\left[f\left(\hp^2_I\right)\delta^i_j+\Bar{f}\left(\hp^2_I\right)\frac{\hp^i_I\hp_{I,j}}{\hp^2_I}\right] \\
    \left[\hx^i_I,\hx^j_J\right] =& 2i\delta_{IJ}\theta (\hp_I^2) \hp_I^{[i}\hx^{j]}_I (\hp^2_I)\,,&\left[\hp_{I,i},\hp_{J,j}\right]=0 \, , \label{GUP}
\end{align}
where $\{i,j\}=\{1,2,3\}$ label vector components whereas $\{I,J\}=\{1,2,\dots,N\}$ denote the $I-$th and $J-$th particle. Given the dynamical variables of the constituents, we may define center-of-mass coordinates and momenta characterising the composite system as
\begin{equation}\label{com}
\hmx^i=\frac{1}{N}\sumi\hx^i_I, \qquad \hmp_i=\sumi\hp_{I,i}.  
\end{equation}
It can be shown that the ensuing deformed commutation relations at the level of center-of-mass positions and momenta read
\begin{align}\label{com2}
\left[\hmx^i,\hmp_j\right]=&\, \frac{1}{N}\sumi\sumj\left[\hx^i_I,\hp_{J,j}\right]=\frac{i}{N}\sumi\left[f\left(\hp^2_I\right)\delta^i_j+\Bar{f}\left(\hp^2_I\right)\frac{\hp^i_I\hp_{I,j}}{\hp^2_I}\right],\\\label{com22}
\left[\hmx^i,\hmx^j\right]=&\, \frac{1}{N^2}\sumi\sumj\left[\hx^i_I,\hx_{J}^j\right]=\frac{2}{N}\sumi\theta(\hp_I^2)\hp_I^{[i}\hx^{j]}_I\, . 
\end{align}
For simplicity, we impose that the collection of constituents undergoes collective quasi-rigid motion. This assumption is particularly tailored to recent experimental proposals aimed at detecting quantum gravitational signatures via quantum optics and mechanical oscillators \cite{Pikovski:2011zk,Bekenstein:2012yy}.
However, it may be violated in more complex systems, for instance in the presence of strong interactions \cite{Kumar:2019bnd}.
The momenta of the constituents of an object in rigid motion are approximately given by $\hp_{I,i}\simeq\hmp_i/N$, $\forall\{I,i\}$.
As a result, the modified commutation relations \eqref{com2} and \eqref{com22} become
\begin{align}\label{com3}
\left[\hmx^i,\hmp_j\right]=&\, i\left[f\left(\frac{\hmp^2}{N^2}\right)\delta^i_j+\bar{f}\left(\frac{\hmp^2}{N^2}\right)\frac{\hmp^i\hmp_j}{\hmp^2}\right],\\
\left[\hmx^i,\hmx^j\right]=&\, \frac{2}{N}\theta\left(\frac{\hmp^2}{N^2}\right)\hmp^{[i}\hmx^{j]}\, .  
\label{com3bis}
\end{align}
Considering the argument of the functions $f,$ $\bar{f}$ and $\theta$, it is evident that the relevance of the deformations decreases quadratically with increasing $N$. Clearly, the thermodynamical limit (\ie, $N\to\infty$) for the above functions is equivalent to the limit $\hmp\to0$, which yields the standard quantum mechanical picture with $f\to1,$ $\bar{f}\to0$ and $\theta/N\to 0$. This implies that macroscopic objects obey ordinary quantum mechanics.

While the GUP thus circumvents the soccer-ball problem, this leaves us with a related issue. In removing the problem of nonlinearity, the reasoning creates an ``inverse'' soccer-ball problem in its place. It is unclear whether the particles deemed as elementary today prove to contain substructure tomorrow. To say that an electron, for example, suffers from Planck-level corrections as introduced here, implies that it marks the final step of the reductionist ladder, a somewhat metaphysical assertion to say the least. This issue appears to sink its roots in the limited nature of single-particle quantum mechanics and is expected to disappear at the level of QFT. Although the above argument does not provide a definite answer to the problem of fundamental constituents, it surely provides an indication towards the settlement of the issue.

\subsection{Heuristic vs. explicit approaches}\label{sec:heurexp}

When browsing the literature on GUPs, it is not uncommon to come across calculations and claims
based on heuristic arguments. Indeed, even though the deviations from standard quantum mechanics are deemed small (and thus treated perturbatively), the computational difficulty might require alternative routes for the solution of the considered problem. Issues of this kind motivate streamlined derivations of complex algebraic outcomes relying on reasonable assumptions that heuristically lead to the same conclusion. A paradigmatic example along this line is given by the GUP-induced corrections to the Casimir effect, which has been originally deduced by means of a full-fledged QFT procedure \cite{Frassino:2011aa}, and later on recovered within a simplified heuristic scheme \cite{Blasone:2019wad,BlasCas2}.

However, heuristic considerations do not always reproduce results stemming from explicit computations. An example of a discrepancy between the two approaches has been recently pointed out in the literature in the context of the Schwinger effect in a (anti-)de Sitter space \cite{Ong:2020tvo}. In this context, it has been demonstrated that, although the corrections to the production rate of particle-antiparticle pairs in the explicit and heuristic computations exactly match in absolute value, there is a fundamental tension between the two approaches in the predicted sign of the contributions. In a nutshell, before the observation contained in \cite{Ong:2020tvo}, it was widely accepted that the presence of a non-vanishing cosmological constant $\Lambda$ can be embedded in nonrelativistic quantum mechanics by means of a deformed commutator between position and momentum operators, namely \cite{Mignemi:2009ji}
\begin{equation}\label{euppo}
    [\hx_i,\hp_j]=i\hbar\delta_{ij}\left(1-\frac{\Lambda}{3}\hx^2\right),    
\end{equation}
with the constant $\Lambda.$ For $\Lambda>0$ ($\Lambda<0$) the model is supposed to mimic a (anti-)de Sitter background. However, for the corrections to the Schwinger effect based on heuristic arguments to be equivalent to explicit calculations, the interpretation of positive (negative) $\Lambda$ as modelling (anti-)de Sitter space has to be reversed.
This contradicts the existing literature on the subject. Although it does not represent a striking rebuttal of heuristic methods, it certainly casts doubt on their general validity.

The previous example is not the only instance that highlights the distinctions between heuristic and explicit results.
As a matter of fact, when considering the GUP-corrected Unruh temperature, there is a mismatch between the heuristic \cite{Nozari:2011gj,Dabrowski:2019wjk,Luciano:2019mrz} and the explicit \cite{Nicolini:2009dr,Majhi:2013koa,Scardigli:2018jlm} predictions.
On the one hand, QFT calculations indicate that GUP corrections, being frequency-dependent, spoil the thermal property of the radiation spectrum.
\footnote{Note that this is not in conflict with the results of \cite{Bisognano:1976za} which state that a thermal spectrum follows from unitarity and relativistic invariance because the latter is violated by the GUP.}
On the other hand, that the thermal spectrum be thermal is a basic assumption of the heuristic approach. Therefore, GUP corrections to the Unruh effect only manifest through a shift of the radiation temperature. The ensuing disparity is not only quantitative but also qualitative.

An analogous conundrum arises in the gravitational context, \ie, when considering the Hawking radiation emitted by black holes: while heuristic derivations \cite{Scardigli:1995qd,Adler:2001vs,Chen:2002tu,Cavaglia:2003qk,Medved:2004yu,Cavaglia:2004jw,Scardigli:2006eb,Park:2007az,Nouicer:2007jg,Custodio:2003jp,Chen:2003bu,Chen:2004ft,Setare:2004sr,Setare:2005sj,Nozari:2005ah,Nozari:2005gn,Zhao:2006ri,Zhao:2007as,Ko:2006mr,Xiang:2006ei,Nozari:2006dq,Nozari:2006rt,Nozari:2006vn,Nozari:2006bi,Zhao:2006xf,Myung:2006qr,Kim:2007hf,Nouicer:2007pu,Arraut:2008hc,Farmany:2009zz,Jizba:2009qf,Xiang:2009yq,Myung:2009va,Dehghani:2009zze,Banerjee:2010sd,Setare:2010ct,Said:2011dg,Majumder:2011xg,Sabri:2012fi,Tawfik:2013uza,Xiang:2013sza,Majumder:2013nia,Gangopadhyay:2013ofa,Awad:2014nma,Scardigli:2014qka,Chen:2014bva,Faizal:2014tea,Tawfik:2015kga,FaragAli:2015boi,NasserTawfik:2015gwy,Gangopadhyay:2015zma,Anacleto:2015rlz,Anacleto:2015awa,Anacleto:2015mma,Anacleto:2015kca,Hammad:2015dka,Dehghani:2015zca,Sakalli:2016mnk,Lambiase:2017adh,Vagenas:2017fwa,Ong:2018xna,Ong:2018syk,Maluf:2018lyu,Maluf:2018ksj,Alonso-Serrano:2018mfo,Alonso-Serrano:2018ycq,Contreras:2018gpl,Buoninfante:2019fwr,Kanazawa:2019llj,Hassanabadi:2019eol,Alonso-Serrano:2020hpb,Petruzziello:2020een,Du:2021hxb,Bosso:2022xnm} \emph{a priori} presume a blackbody spectrum (thus only modifying the value of the Hawking temperature), explicit ones \cite{Nozari:2008gp,Majumder:2012rtc,Nozari:2012nf,Chen:2013pra,Chen:2013tha,Chen:2013ssa,Feng:2015jlj,Ovgun:2015jna,Sakalli:2016mnk,Li:2016mwq,Ovgun:2017hje,Gecim:2017nbh,Gecim:2017zid,Kanzi:2019gtu} based on the Parikh-Wilczek \cite{Parikh:1999mf} formalism  lead to frequency- and mass-dependent corrections.
This is also implied by the calculations on the Unruh effect.
\footnote{To see how the Hawking effect may be understood as an instance of the Unruh effect, see \eg \cite{Padmanabhan:2009vy}.}
The non-thermal behavior has been argued to possibly solve the information paradox by letting information leak out \cite{Nozari:2008gp}. This highlights that there can be a qualitative difference between heuristic and explicit ans\"atze. 

Notwithstanding the drawbacks of heuristic reasoning, in the GUP framework there are problems to which explicit solutions remain elusive. Such is the case, for example, for minimal-length modifications to the gravitational field equations. While there are approaches to GUP effects in QFT (see Sec. \ref{Rext}), the non-linear nature of the full gravitational field equations harbours additional subtleties, as of yet preventing their deformation. In its place, there are a number of heuristic approaches towards a minimal-length modification of the Schwarzschild black hole metric, namely an emergent-gravity scenario based on the GUP modified black hole entropy \cite{FaragAli:2015boi,Contreras:2016xib,Casadio:2020rsj}, a deformation of minisuperspace models in canonical quantum gravity \cite{Bina:2010ir,Majumder:2011bv,Bosso:2019ljf} and a reparametrization of the ADM mass which, for small masses, mimics the Compton wavelength \cite{Carr:2011pr,Carr:2014mya,Buoninfante:2020cqz}. The proliferation of those different results further highlights the subtleties heuristic approaches entail, as reviewed in \cite{Ong:2023jkp}.

In a nutshell, we do not suggest to \emph{a priori} reject results achieved by means of heuristic arguments; instead, we are warning to instill great care when applying this kind of analysis. When allowed by the mathematical complexity of the considered problem, it is preferable to rely on the explicit approach, and apply heuristic methods solely to gain an intuition.

For example, many of the bounds on the GUP parameter are based on heuristic reasoning and require further assessment. These constraints can be found in the next subsection.

\subsection{Bounds on the model parameter}
\label{boun}
As a phenomenological model, much of the literature on GUPs naturally consists in constraints on model parameters. However, not all bounds were obtained applying equal standards of rigour. Here, we try to give a comprehensive classification of the existing constraints in the literature (see Tabs. \ref{tab:labgup}, \ref{tab:gravigup} and \ref{tab:cosmogup}). Over all, there are four different ways along the lines of which these kinds of studies have proceeded:
\begin{itemize}
    \item Direct application of the deformed commutation relations to quantum mechanical systems. As such, they allow for the clearest interpretation. These studies, colour-coded green in the tables, mainly concentrate on tabletop experiments (Tab. \ref{tab:labgup}).
      As for evaluations based on analyses related to the hydrogen atom, while there have been claims in the literature that the corrections are nonperturbative leading to stronger bounds \cite{Fityo:2005xaa,Bouaziz:2010he,Pedram:2012ub}, in the respective works the absolute value of the radius appearing in the Coulomb potential was neglected, thus solving for positive and negative radii. Therefore, their findings are still under debate. They have been refuted in \cite{Slawny:2007pya}, which has been corroborated by numerical results in \cite{Benczik:2005bh,Chang:2011jj}.
    \item Modifications to the black-hole and cosmological spacetimes derived from heuristic corrections to the Hawking temperature in the context of emergent gravity. As explained in Sec. \ref{sec:heurexp}, the interpretation of these derivations is still under debate. Indicated in orange, they naturally deal with astrophysical (Tab. \ref{tab:gravigup}) and cosmological (Tab. \ref{tab:cosmogup}) observations.
    \item Deformed Poisson brackets as classical counterparts of GUPs. For the reasons outlined in Sec. \ref{sec:conslim}, those are slightly more subtle to interpret, particularly in the case of applications involving the gravitational field. They are indicated in yellow in Tab. \ref{tab:cosmogup}.
    \item A redefinition of the ADM-mass such that it mimics the mass-dependence of the Compton wavelength at \mbox{(sub-)Planckian} scales in accordance with the black hole uncertainty principle correspondence \cite{Carr:2011pr,Carr:2014mya,Carr:2015nqa}. As a reparametrization of a free constant in terms of another one, it is unclear whether the corrections to gravitational observables obtained using this approach are physical. Finding their natural arena in astrophysical observations (Tab. \ref{tab:gravigup}), these studies are indicated in red.
\end{itemize}
The careful reader may have realized that Tabs.~\ref{tab:labgup} and \ref{tab:gravigup} contain entries with gray background in addition to the coloured ones. These indicate studies based on deformations the Heisenberg algebra for macroscopic objects, whose (macroscopic) mass serves as an amplifier. 
Unfortunately, as laid out in Sec. \ref{sec:FundConst} and \cite{Quesne:2009vc,Tkachuk:2012gyq,Amelino-Camelia:2013fxa}, this is problematic: the modifications to the canonical commutator should scale with the squared inverse number of constituents.
Correcting for this issue, the bounds on $\beta$ weaken by a factor of at least $10^{44},$ thereby limiting the usefulness of macroscopic bodies to the search for a minimal length.

In a nutshell, we see that the most rigorous bound on the model parameter of the quadratic GUP stems from precision measurements of the 1S-2S transition frequency of the hydrogen atom in the context of LIV \cite{Gomes:2022hva}, reading
\begin{equation}
    \beta <10^{30}.
\end{equation}
It is worth observing that the above bound  corresponds to the scales accessed at the LHC \cite{Evans:2008zzb}.
The road to Planckian precision remains long, and the field is in dire need for new ideas to make progress.
Here, the synthesis of the GUP and other approaches in quantum gravity phenomenology could be of use. For instance, the transfer of constraints on MDRs from time-of-flight measurements implies the bound
\begin{equation}
    \beta\leq 10^{16},
\end{equation}
thus providing a gain of fourteen further orders of magnitude.

\begin{table}[!ht]
    \centering
\begin{tabular}{ c c || c} 
\toprule
 experiment & ref.  & upper bound on $\beta$ \\
 \midrule
 \rowcolor{gray!40}phonon cavity & \cite{Campbell:2023dwm}& \sout{$62$} $10^{46}$\\ \rowcolor{gray!40}harmonic oscillators & \cite{Bushev:2019zvw,Bawaj:2014cda} & \sout{$10^7$} $10^{60}$\\ 
 \rowcolor{nicegreen!40}{LIV in hydrogen atom} & \cite{Gomes:2022hva} & $10^{30}$\\
  \rowcolor{nicegreen!40}scanning tunnelling microscope & \cite{Pikovski:2011zk,Das:2008kaa} & $10^{33}$\\
 \rowcolor{nicegreen!40}$\mu$ anomalous magnetic moment & \cite{Das:2011tq,Das:2008kaa} & $10^{33}$\\
 \rowcolor{nicegreen!40}Hydrogen atom
 & \cite{Benczik:2005bh,Slawny:2007pya,Chang:2011jj,AntonacciOakes:2013qvs,Brau:1999uv} & $10^{34}$\\
 \rowcolor{nicegreen!40}lamb shift & \cite{Das:2008kaa,Ali:2011fa} & $10^{36}$\\ 
 \rowcolor{nicegreen!40}$^{87}$Rb interferometry & \cite{Gao:2016fmk,Khodadi:2018kqp} & $10^{39}$\\
 \rowcolor{nicegreen!40}Kratzer potential & \cite{Bouaziz:2013ora} & $10^{46}$\\
 \rowcolor{nicegreen!40}stimulated emission    &   \cite{Twagirayezu:2020chx}    &   $10^{46}$\\
 \rowcolor{nicegreen!40}Landau levels & \cite{Ali:2011fa,Das:2008kaa,Das:2009hs} & $10^{50}$\\
 \rowcolor{nicegreen!40}quantum noise & \cite{Girdhar:2020kfl} & $10^{57}$\\
 \bottomrule
\end{tabular}
\caption{Upper bounds on the quadratic GUP-parameter by tabletop experiments not related to gravity. Tests which are rigorously relatable to modified commutators are marked in green, experiments involving macroscopic quantum objects in gray. \label{tab:labgup}}
\end{table}

\begin{table}[!ht]
    \centering
\begin{tabular}{ c c || c} 
\toprule
 experiment & ref.  & upper bound on $\beta$ \\
 \midrule
 \rowcolor{orange!40}perihelion precession (solar system, 1) & \cite{Benczik:2002tt,Quesne:2009vc} &\sout{$10^{-68}$} $10^{34}$\\
 \rowcolor{yellow!40}{time-of-flight measurements} & \cite{Wagner:2023fmb}& $10^{16}$\\
 \rowcolor{gray!40}equivalence principle (pendula) & \cite{Ghosh:2013qra} & \sout{$10^{20}$} $10^{73}$\\ 
 \rowcolor{gray!40}gravitational bar detectors & \cite{Marin:2013pga,Marin:2014wja} & \sout{$10^{33}$} $10^{93}$\\
\rowcolor{nicegreen!40}equivalence principle (atoms) & \cite{Gao:2017zch} & $10^{45}$\\
\rowcolor{yellow!40}low-mass stars & \cite{Pachol:2023bkv} &$10^{48}$\\
 \rowcolor{orange!40}{LIV in torsion pendulum} & \cite{Lambiase:2017adh} &$10^{51}$\\
 \rowcolor{orange!40}perihelion precession (solar system, 2) & \cite{Scardigli:2014qka,FaragAli:2015boi} &$10^{69}$\\
 \rowcolor{orange!40}perihelion precession (pulsars) & \cite{Scardigli:2014qka} &$10^{71}$\\
 \rowcolor{orange!40}gravitational redshift & \cite{FaragAli:2015boi} & $10^{76}$\\
 \rowcolor{red!40}black hole quasi normal modes & \cite{Jusufi:2020wmp} & $10^{77}$\\
 \rowcolor{orange!40}light deflection & \cite{Scardigli:2014qka,FaragAli:2015boi} & $10^{78}$\\
 \rowcolor{orange!40}time delay of light & \cite{FaragAli:2015boi} & $10^{81}$\\
 \rowcolor{orange!40}black hole shadow & \cite{Neves:2019lio} & $10^{90}$\\
  \rowcolor{red!40}black hole shadow & \cite{Jusufi:2020wmp,Tamburini:2021inp} & $10^{90}$\\
  \bottomrule
\end{tabular}
\caption{Upper bounds on the quadratic GUP-parameter by gravitational experiments and observations. Tests which are rigorously relatable to modified commutators are marked in green, experiments involving macroscopic quantum objects in gray, while results with orange background are based on the heuristic application of the GUP to gravitational thermodynamics. The red rows contain instances of the black-hole mass-modification approach, while the yellow background stands for Poisson-bracket descriptions of the GUP.
\label{tab:gravigup}}
\end{table}

\begin{table}[!ht]
    \centering
\begin{tabular}{ c c || c}
\toprule
 experiment & ref.  & upper bound on $\beta$ \\
 \midrule
 \rowcolor{yellow!40}gravitational waves & \cite{Das:2021lrb,Feng:2016tyt} & $10^{36}$\\
 \rowcolor{orange!40}cosmology (all data) & \cite{Giardino:2020myz} & $10^{59}$\\
 \rowcolor{orange!40}cosmology (late-time) & \cite{Giardino:2020myz,Kouwn:2018rmp} & $10^{81}$\\
 \bottomrule
\end{tabular}
\caption{Upper bounds on the quadratic GUP-parameter on cosmological scales. Results with orange background are based on the heuristic application of the GUP to gravitational thermodynamics, while the yellow background stands for Poisson-bracket descriptions of the GUP. \label{tab:cosmogup}}
\end{table}

\section{Discussion and Conclusions}
\label{DeC}

In this review, we have critically discussed some shortcomings and open problems arising within the framework of GUPs, 
which are often overlooked or naively addressed 
in the pertinent literature.
We summarize our main results and conclusions below:
\begin{itemize}
\item Besides formal similarities, relativistic corrections to Heisenberg's uncertainty relation and the GUP are intrinsically different. In fact, while the former lead to a particle-specific deformed Hamiltonian, the latter are expected to be universal. In this context, we have also discussed consistent limits of the GUP, showing that, contrary to recent statements in the literature, it can be well-motivated to analyze corrections to classical dynamics
from Planck-scale suppressed effects, although the
interpretation may be subtler than in the quantum regime (see Sec.~\ref{RvG} and~\ref{sec:conslim}).

\item Reconsidering the essence of the minimal length, it becomes clear that the customary choice of Hamiltonian is not justified by the existence of the minimal length alone; instead, it introduces additional structure. This raises the question: which choice of Hamiltonian is preferable? In order to make an informed decision on the matter, a detailed symmetry analysis is required which is, as of yet, missing. Details can be found in Sec. \ref{sec:SymHam}. 

\item Even though it is frequently adopted in the literature, the na\"ive relativistic extension~\eqref{RelHeisAlg} of the Heisenberg algebra to Lorentzian spacetime is vitiated by some conceptual subtleties, such as the impossibility of defining a self-adjoint time operator 
for physical systems with unbounded energy and the fact that position is no longer an operator in a relativistic QFT. These motivations severely question the meaning of relativistic uncertainty relations of Heisenberg-type (details are contained in Sec.~\ref{Rext}).

\item A consistent relativistic generalization of the Heisenberg algebra requires deformations of QFT, for example to describe the standard model. Despite the number of approaches introduced so far, 
a generally accepted solution is still lacking. The discussion on this point has been presented in Sec.~\ref{Rext}. 

\item Quantum-gravitational effects deteriorate as the number of elementary constituents of a given composite system increases. This implies a nontrivial rescaling of GUP corrections. On the one hand, this renders the detection of quantum-gravitational signatures in macroscopic objects challenging. On the other hand, it raises a conceptual question: what are the fundamental degrees of freedom? As it requires information on the content of the universe at all, including unknown scales, assumptions on the matter have a rather metaphysical flavour. This issue (a sort of ``inverse'' soccer ball problem) has been discussed in Sec.~\ref{sec:FundConst}.

\item Although heuristic considerations within the GUP framework can compensate for computational technicalities, they might be misleading. We have provided several examples for which the ensuing results contradict more rigorous and explicit arguments. This mismatch should not discourage the usage of heuristic reasoning in general, because it frequently helps to gain an intuition on the nature of Planck-scale effects. Rather, the examples are to be taken as a warning not to recline on heuristics, but possibly to use them as a starting point towards reaching the correct solution (see Sec.~\ref{sec:heurexp} for more details).

\item In the 30 years of its existence, the field of GUPs has amassed a great number of constraints on the model parameter. We have presented and classified a comprehensive collection of existing bounds in Tabs. \ref{tab:labgup},~\ref{tab:gravigup}, and~\ref{tab:cosmogup}. In this context, it is noteworthy that there are applications of the GUP to macroscopic objects which do not take into consideration that Planck-scale effects deteriorate with increasing number of degrees of freedom. The most stringent constraint amounts to $\beta<10^{30}$,  which indicates that there is still much room for improvement on the experimental side of GUP framework (see Sec.~\ref{boun}).
\end{itemize}
30 years into the research on GUPs, even foundational matters such as the inverse soccer ball problem are not entirely resolved. More effort is required for the construction of a well-posed Planck-scale deformation of the Heisenberg uncertainty principle. Rather than just pointing out difficulties, this review is intended to inspire new research directions towards a rigorous understanding of minimal-length phenomenology.

\section*{Acknowledgements}
The authors acknowledge networking support by the COST Action CA18108.  G.G.L. is grateful to the Spanish ``Ministerio de Universidades'' for the awarded Maria Zambrano fellowship and funding received
from the European Union - NextGenerationEU. L.P. acknowledges support by MUR (Ministero dell’Universit\`a e della Ricerca) via the project PRIN 2017 ``Taming complexity via QUantum Strategies: a Hybrid Integrated Photonic approach'' (QUSHIP) Id. 2017SRNBRK and is grateful to the ``Angelo Della Riccia'' foundation for the awarded fellowship received to support the study at Universit\"at Ulm.

\section*{References}


\providecommand{\href}[2]{#2}\begingroup\raggedright\endgroup
\end{document}